# FAST ELECTRONICS FOR THE DAΦNE TRANSVERSE FEEDBACK SYSTEMS

A. Drago, A. Argan, M. Serio,
Laboratori Nazionali di Frascati - INFN, Frascati, Italy

*Abstract*

Transverse feedback systems for controlling the vertical coupled-bunch instabilities in the positron and electron main rings are installed at DAΦNE. They started to be operative respectively from June and September 2000. For the horizontal plane, similar systems have been installed in summer 2001 with less kicker power. Design specifications and the basic system concepts are presented. Real time bunch-by-bunch offset correction is implemented using digital signal processors and dual-port RAM's. Fast analog to digital sampling is performed at the maximum bunch frequency (368 MHz). The system manages at full speed a continuous flow of 8-bits data and it has the capability to invert the sign or put to zero the output for any combination of bunches. A conversion from digital to analog produces the output correcting signal.

## 1 INTRODUCTION

DAΦNE is a φ−factory, mainly dedicated to the study of CP violation, currently in operation at Frascati (Italy). It has achieved a peak luminosity of ~$5*10^{31}$ $cm^{-2}s^{-1}$ in the Kloe detector interaction region. Moreover, maximum currents of 1411mA in the electron ring and 1150mA in the positron one have been stored. In Table 1, some DAΦNE parameters are shown.

Table 1: DAΦNE parameters

| Ring | e- / e+ |
|---|---|
| Energy | 0.510 GeV |
| Circumference | 97 m |
| RF frequency | 368.29 MHz |
| Harmonic # | 120 |
| Rev. frequency | 3.069 MHz |
| Betatron tune x | 5.1170 / 5.1615 |
| Betatron tune y | 5.1623 / 5.2244 |
| Horizontal freq. | 360 / 495 kHz |
| Vertical freq. | 498 / 689 kHz |
| Max bunch curr | ≤44.1 mA |
| Bunch distance | 2.7 nsec |
| Typ. fill pattern | 47 stepped by 2 |

At the beginning of 2000, during commissioning, it was decided to install vertical feedback systems on the two main rings. Similar devices have been successfully used in other colliders or synchrotron light sources [1 - 7], however the design of our system is peculiar under some aspects. The design of the system developed at the Frascati laboratory was stepped in two phases: the first one was made operative in June 2000 and at the begin of 2001 the power of the back end stage was increased [8]. The e+ system is shown in figure 1.

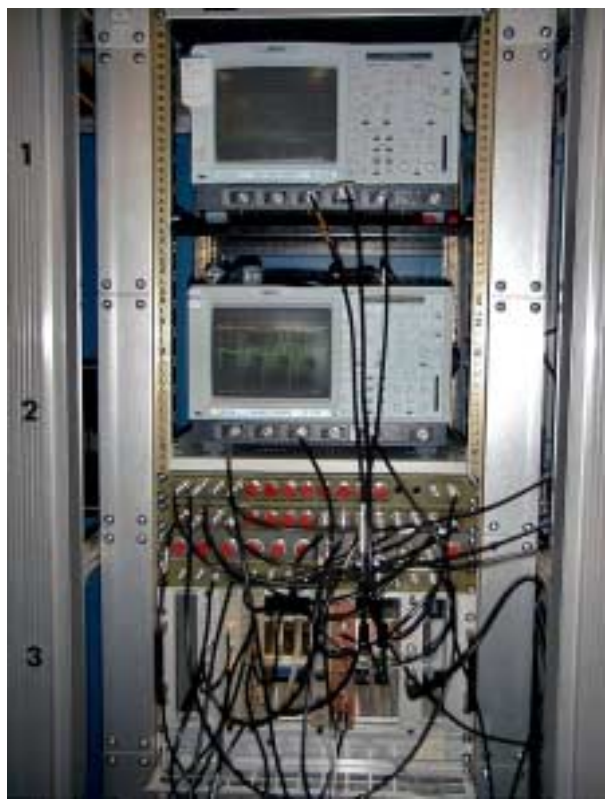

Figure 1: Vertical Feedback running for the positron ring.

In the second phase, the feedback system includes new modules added to the first version setup; it is currently under test. It can allow more flexibility in the system and a better performance of the power stage by computing a more efficient signal by a bunch-by-bunch offset correction scheme.

## 2 THE INSTALLED TRANSVERSE FEEDBACK SYSTEM

A four buttons beam position monitor produces pulses at the passage of a bunch. It is possible to generate a difference signal connecting the up and down (or left and right) button to a hybrid junction. Fast analog to digital sampling of the signal is performed at base band at the DAΦNE Radio Frequency (368 MHz) by a MAX101A, an 8-bits ADC by MAXIM. The system manages at full speed a continuous stream of data and it has the capability to invert the sign or put to zero the output for single or multiple values in real time. To manage a so fast flow of data, ECL components from the ECLinPS family MC100E (by Motorola) are used. Printed circuit boards with traces at controlled impedance have been designed at the Frascati laboratory. The skew between 16+1 differential signals is a very critical parameter and requires a careful lay out of the PCB traces. A conversion from digital to analog produces the bunch-by-bunch error signal that, after a stage of pre-amplification, is sent as a correcting signal to the power amplifiers. Figure 2 shows the correction signal and the kicker signal. In this implementation of the system local orbit bumps are used to avoid saturation of the power amplifiers due to the static and not useful part of the signal coming from the orbit.

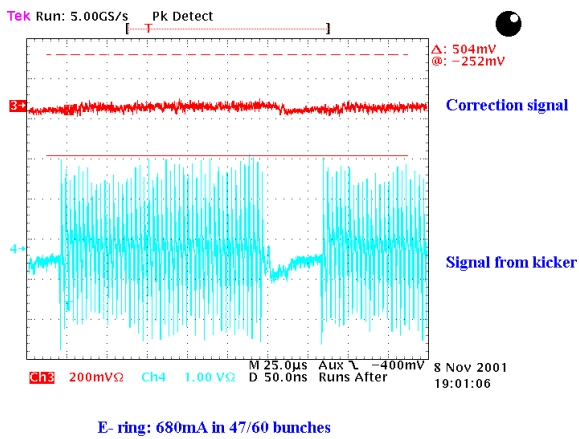

Figure 2: Feedback output signals.

## 3 THE UPGRADED FEEDBACK SYSTEM

The upgraded version of the system adds two modules to the previous ones to improve flexibility and efficiency of the power stage. The simplified scheme is shown in Figure 3.

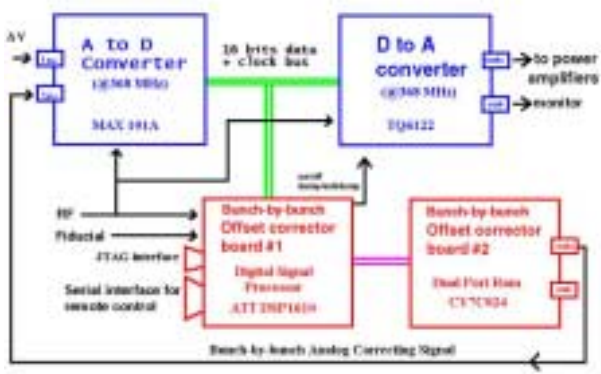

Figure 3: Scheme of the complete feedback system.

A Digital Signal Processor ATT DSP1610 can record a block of data relative to a selected bucket at the revolution frequency (>3MHz). In a typical run, the DSP computes and updates continuously the average value for each bucket (empty or not). Then it puts the results in a dual port RAM (CY7C024 by CYPRESS). An 8 bits digital to analog converter (TQ6122 by Triquint) outputs at 368MHz the value to be subtracted from the input signal. The hold buffer is circular and it is scanned with a proper delay. The four modules are shown in figure 4.

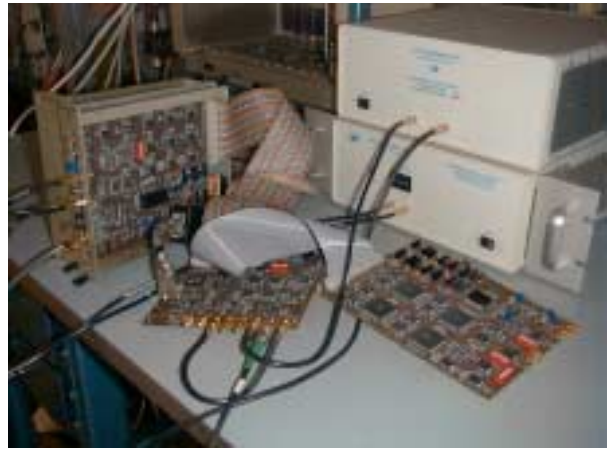

Figure 4: The complete feedback system in the lab.

The digital signal processor can address up to 128 Kbytes of programs, including an FFT routine. It has a JTAG interface for debugging purposes. Moreover, through a 4 MHz synchronous serial interface that can connect up to eight slave devices to a master, it is possible to exchange commands and data following the pseudo language listed in Table 2. The post processing software still is under development.

Table 2: Multipoints serial interface commands list

| | |
|---|---|
| FO | Feedback on |
| FF | Feedback off |
| TE | Test |
| DA | Damping feedback |
| AD | Antidamping feedback |
| OC | Start offset corrector |
| ST | Stop offset corrector |
| Rxx | Reduce correction at xx% |
| Nxx | Number_of_average/100 [0:255], i.e. from zero to 25500 |
| Bxx | Bunch selected [1:120], 0=scan forever on all the bunches |
| I | Store a block of data for a selected bunch to internal RAM |
| J | Download a block of data from internal RAM to serial interface |
| M | Store a block of data for a selected bunch to dual port RAM |
| L | Download a block of data from dual port RAM to serial interface |
| Sxx | Select slave device (0:7) |
| Txx | Compute tune bunch xx (1:120, 0=all) |

The system shown in figure 4 is currently under test in the laboratory. A good manner to evaluate its linearity and to prove that the internal skew is adequate can be realized by sending to the input a ramp in the working range of the ADC. Then the digital data cross the modules and at the end of the chain are converted again to an analog signal. The result of this test is shown in figure 5.

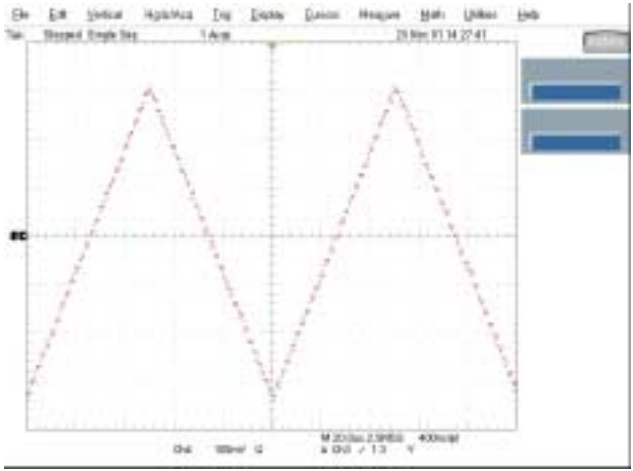

Figure 5: Analog output of the test signal.

## 4 CONCLUSIONS

The vertical feedback system damps efficiently coherent instabilities in DAΦNE, allowing the storage of 1411 mA in the electron ring and 1150 mA in the positron one. The increase of the stored currents was a crucial item in the improvement of peak and average luminosity. The system has shown a reliable behaviour during more then one year and it has offered in the same time useful diagnostics to manage it correctly. The next upgrade of the system is expected to further improve DAΦNE performance.

## 5 ACKNOWLEDMENTS

Donato Pellegrini, Olimpio Giacinti, Gianfranco Baccarelli and Giuseppe Fallica under the supervision of Oscar Coiro have taken care of system setups and cabling. Umberto Frasacco has skillfully loaded the boards. F. Ronci and F. Galletti have prepared the printed circuit layouts. Thanks to Pina Possanza for the patient text editing.

Thanks also to Marco Lonza and Daniele Bulfone for ideas and data exchanges on the power amplifiers and to John Fox, Dmitry Teytelman and Shyam Prabhakar for many interesting conversations.